\documentclass[a4paper,aps,twocolumn,prb,
showpacs,amsmath,amssymb,superscriptaddress]{revtex4-1}

\usepackage{graphicx}
\usepackage{dcolumn}
\usepackage{bm}
\usepackage{longtable}
\usepackage{color}

\makeatletter
\newcommand\figcaption{\def\@captype{figure}\caption}
\newcommand\tabcaption{\def\@captype{table}\caption}

\makeatother

\begin{document}


\title{Dimensionality-tuned electronic structure of nickelate superlattices explored by soft-x-ray angle resolved photoelectron spectroscopy}

\author{G. Berner}
\author{M. Sing}
\author{F. Pfaff}
\affiliation{Physikalisches Institut and R\"ontgen Center for Complex Material Systems (RCCM), Universit\"at W\"urzburg, Am Hubland,
D-97074 W\"urzburg, Germany}
\author{E. Benckiser}
\author{M. Wu}
\author{G. Christiani}
\author{G. Logvenov}
\author{H.-U. Habermeier}
\affiliation{Max Planck Institute for Solid State Research, Heisenbergstrasse 1, 
D-70569 Stuttgart, Germany}
\author{M. Kobayashi}
\author{V.N. Strocov}
\author{T. Schmitt}
\affiliation{Swiss Light Source, Paul Scherrer Institut, CH-5232 Villigen, Switzerland}
\author{H. Fujiwara}
\affiliation{Division of Materials Physics, Graduate School of Engineering Science, Osaka University, Osaka 560-8531, Japan}
\author{S. Suga}
\author{A. Sekiyama}
\affiliation{Institute of Scientific \& Industrial Research, Osaka University, Ibaraki, Osaka 567-0047, Japan}
\author{B. Keimer}
\affiliation{Max Planck Institute for Solid State Research, Heisenbergstrasse 
1, D-70569 Stuttgart, Germany}
\author{R. Claessen}
\affiliation{Physikalisches Institut and R\"ontgen Center for Complex Material Systems (RCCM), Universit\"at W\"urzburg, Am Hubland,
D-97074 W\"urzburg, Germany}

\date{\today}

\begin{abstract}

The electronic and magnetic properties of epitaxial LaNiO$_3$/LaAlO$_3$
superlattices can be tuned by layer thickness and substrate-induced
strain. Here, we report on direct measurements of the $k$-space-resolved
electronic structure of \textit{buried} nickelate layers in superlattices under 
compressive strain by soft x-ray photoemission. After disentangling strong
extrinsic contributions to the angle-dependent signal caused by photoelectron
diffraction, we are able to extract Fermi surface information from our data. We
find that with decreasing LaNiO$_3$ thickness down to two unit cells (2 uc) 
quasiparticle
coherence becomes strongly reduced, in accord with the dimension-induced 
metal-to-insulator transition seen in transport measurements. Nonetheless, on top 
of a strongly incoherent background a residual Fermi surface can be identified in 
the 2 uc superlattice whose nesting properties are consistent with the spin-density 
wave (SDW) instability recently reported. The overall behavior of the 
Ni~$3d$ spectra and the absence of a complete gap opening indicate that the SDW 
phase is dominated by strong order parameter fluctuations. 
\end{abstract}

\pacs{79.60.-i, 79.60.Jv, 73.20.-r, 73.50.Pz}

%
%

\maketitle

\section{Introduction}

Artificial heterostructures made from transition metal oxides may host novel
electronic and magnetic phases not present in the bulk of the constituents. Such phases may be controlled by, {\it e.g.}, elastic strain or interfacial charge
transfer.\cite{Hwang12} In this context the perovskite LaNiO$_3$ (LNO) is a very
interesting material, as it is the only member of the $RE$NiO$_3$ ($RE =$ rare
earth, \textit{viz.}, La, Pr, Nd, Sm, Eu) family showing metallic behavior with its
partially-filled degenerate Ni~$3d$ $e_{g}$ ($d_{z^2}$ and $d_{x^2-y^2}$)
orbitals, while the other compounds exhibit a correlation-driven 
metal-insulator (MI) transition at low temperatures.\cite{Torrance92, Imada98} 
LNO is thus a highly correlated metal being close to an insulating phase, 
which may be switchable by interfacing.

Indeed, recent investigations have shown that the physical properties of LNO can
be tuned by reducing its dimensionality or applying strain. For instance,
density functional theory (DFT) calculations found that a single unit cell (uc)
of LNO sandwiched between layers of the band insulator LaAlO$_3$ (LAO) and thus
under tensile strain displays a cuprate-like Fermi surface,
identifying LNO as a possible candidate for exotic
superconductivity.\cite{Chaloupka08, Hansmann09, Hansmann10} Inspired by these
predictions a large number of experimental and theoretical studies have
been performed on ultra-thin films and superlattices (SLs). \cite{Scherwitzl09,
Scherwitzl11, Sakai13, Yoo13, Benckiser11, Boris11, Chakhalian11, Kaiser11,
Gray11, Freeland11, Frano13, Wu13, Han11, Lee11, Lee11a}

Experimentally, ultra-thin films are found to exhibit a dimensional crossover
from metallic to insulating behavior upon reducing the film thickness, with
critical values ranging from 3 to 9 uc \cite{Scherwitzl09,
Scherwitzl11, Sakai13, Yoo13}. The situation is slightly different in LNO-based
SLs: while thicker LNO layers embedded in SL structures remain metallic at all 
temperatures, 2 uc layers display a temperature-dependent 
MI transition.\cite{Boris11} The critical temperature depends mainly on the
epitaxial strain induced by the substrate. The strain also leads to a distortion
of the octahedral ligand field and hence a lifting of the $e_g$ orbital
degeneracy. The resulting orbital polarization increases almost linearly with
the induced strain. \cite{Frano13, Wu13}. The reason for the appearance of the
insulating state in both SLs and ultra-thin films is still under debate.
Different scenarios like Anderson localization, \cite{Scherwitzl09,
Scherwitzl11} Mott insulator transition, \cite{Hansmann09, Hansmann10} charge
disproportionation, \cite{Chakhalian11, Liu11} and spin density wave
\cite{Boris11, Frano13} have been discussed.

It should be noted that in superlattices the necessity of octahedral 
connectivity across the interfaces in combination with strain induced by the 
substrate can stabilize different distortions compared to the situation in 
epitaxial films.\cite{Hwang13,May10} We emphasize in particular that in 
comparison to ultrathin films the superlattice structures
allow for a better defined and efficient control of the strain effects. Since LNO
with its LaO$^+$ and NiO$_2^-$ sublayers is polar, the (001)-oriented
free surfaces of thin LNO films tend to reconstruct structurally by polar
distortions or octahedral rotations to minimize the electrostatic energy.
\cite{May10, May11, Kumah14} The modified structure results in a change of the
electronic structure, at least near the surface. In contrast, the
LNO layers in the SLs are embedded in LAO, a host material of like polarity, 
and are thus stabilized against structural effects resulting from polar discontinuities, as shown by x-ray 
diffraction.\cite{Kumah14} 

Thus, the physical properties of the strained LNO layers and their 
dependence on dimensionality can be studied under clean conditions, \textit{i.e.}, 
in the absence of additional ionic or electronic surface and interface reconstructions

%
%
\begin{figure*}
\includegraphics[width = 0.96\textwidth]{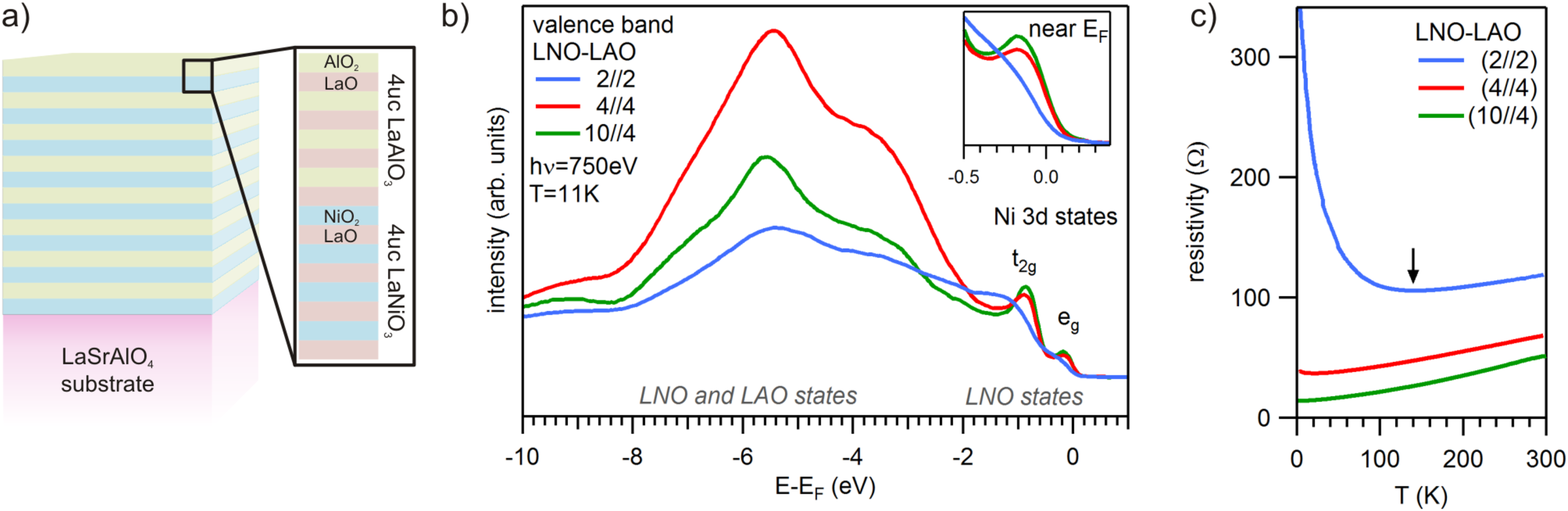}
\caption{\label{Fig:AngleIntSpec} (Color online) (a) Layout of the (4//4)
LNO/LAO superlattice. The first LNO layer is buried below four uc LAO. (b)
$k$-integrated valence band spectra of the investigated LNO/LAO samples at low
temperature. The (2//2) SL exhibits a dramatic loss of quasiparticle coherence
at the Fermi level (see inset). The spectra are normalized to the same
integrated area between $-2$~eV and $E_F$. (c) Temperature-dependent
resistivity measurements. The (2//2) SL shows a continuous MI transition at $T
\approx 150$~K, while the (4//4) and (10//4) SLs stay metallic down to very low
temperatures.}
\end{figure*} 

Direct insight into the electronic structure can be achieved by photoelectron
spectroscopy (PES). Many of the scenarios mentioned above are reflected directly
or indirectly in the microscopic electronic structure, {\it i.e.}, in the
single-particle spectral function, to which PES provides direct access. Such
investigations have focused so far on (ultra-thin) films by 
angle-integrated \cite{Sakai13} as well as angle-resolved PES (ARPES).
\cite{Eguchi09, Yoo13, King14} However, ARPES studies on SLs require a more
bulk-sensitive approach, since the LNO layers are buried below several unit
cells of the host material (LAO in our case). Conventional photoemission with
low-energy photons is limited by its very low information depth of a few
Angstroms only, determined by the photoelectron mean free path (MFP) 
and therefore cannot easily access the LNO layers. In contrast, photons in
the soft x-ray (SX) regime allow for higher MFP at still reasonable 
momentum resolution ($k$-resolution), thereby enabling $k$-resolved probing of 
the buried electronic structure in such SLs.\cite{Berner13,Strocov14}

In this study, we present SX-ARPES measurements on LNO/LAO SLs under compressive
strain and a detailed analysis of the buried electronic structure. Besides a
significant loss of quasiparticle (QP) coherence near the Fermi level ($E_F$)
for the two-dimensional ground state, observed in the angle-integrated valence
band spectra, our $k$-resolved measurements trace the dimensional crossover from
a three-dimensional Fermi surface in the 4 uc LNO-SL to two-dimensional
behavior in the 2 uc LNO-SL. Although the intensity distribution in our
$k$-space maps is strongly affected by x-ray photoelectron diffraction (XPD), we
are able to extract bandstructure information on the Ni $d_{x^2-y^2}$-derived hole pocket states in
all LNO/LAO SLs. An indication of the Ni $d_{z^2}$-derived electron pocket is
only observed in the 2 uc LNO-SL. The residual Fermi surface observed in this
sample displays strong nesting properties which are consistent with the
spin-density wave (SDW) scenario recently put forward as explanation for the
insulating low-temperature ground state.

\section{Experiment}

LNO/LAO superlattices with different LNO layer thickness were grown by pulsed
laser deposition on (001) LaSrAlO$_4$ (LSAO) single crystal substrates. The
deposition starts with a number of $N$ (=2, 4) unit cells of LNO
followed by the same number of LAO layers. This stacking sequence ($N$//$N$) is
repeated fifteen and eight times for $N$=2, 4, respectively, terminating with an LAO layer
at the surface [see Fig. \ref{Fig:AngleIntSpec} (a)]. Additionally, a
(10//4)$\times 4$ SL was prepared being used as a thick LNO reference layer,
also capped by 4 uc LAO.

The LNO/LAO stacks are compressively strained to the LSAO substrate, since the
lattice mismatch between the LSAO and bulk pseudo-cubic LNO ($a=3.838$~{\AA})
\cite{Garcia-Munoz92} is $\approx -3.2$\% (Ref.~\onlinecite{Wu13}). The induced
biaxial strain, which is stable up to a SL thickness of $\approx 50$~nm
(Ref.~\onlinecite{Frano14}), was confirmed by x-ray diffraction with
lattice parameters $a = 3.750$~{\AA} and $c = 3.840$~{\AA}
(Ref.~\onlinecite{Wu13}).

The photoemission experiments were performed at the ADRESS beamline of the Swiss
Light Source with a SPECS Phoibos 150 spectrometer using $p$-polarized
photons.\cite{Strocov10, Strocov14} The overall energy resolution was 
$70$~meV at a photon energy of 700~eV. During the experiment the samples were
cooled down to $T = 11$~K. Preliminary SX-ARPES experiments were performed
at beamline BL25SU, SPring-8, Japan, to establish the feasibility of the method.
Prior to the measurements the sample surface was cleaned by keeping the samples
under ozone flow for 45 min, followed by an {\it in situ} annealing at
180$^\circ$C under $1\times10^{-5}$~mbar of oxygen for 45 min. This method was
found to strongly suppress the amount of carbon-containing surface contaminants.

\section{Results}

\subsection{Angle integrated spectra}

Figure~\ref{Fig:AngleIntSpec}~(b) shows the angle integrated valence band (VB)
spectra of the (2//2), (4//4), and (10//4) SLs measured in normal emission
geometry at low temperature ($T=11$ K). Below $-2$~eV the valence band mainly 
consists of
superposed LAO and LNO O~$2p$-derived states. Due to the valence band offset of
$\approx 2$~eV between LAO and LNO (see Appendix~\ref{Sec:App1}) the spectral
weight between $-2$~eV and $E_F$ can unambiguously be assigned to LNO states, in particular to
the crystal-field split Ni~$3d$ $t_{2g}$ and e$_{g}$ states at $\approx -0.8$~eV
and at $E_F$, respectively. The octahedral splitting is consistent with 
values observed in thin LNO films. \cite{Horiba07, Gray11, Sakai13} The spectra are
normalized to the same integrated area between $-2$~eV and $E_F$ in order to 
facilitate 
easy identification of the changes in the Ni 3d states in this energy window. A normalization 
to the integral intensity of the \textit{full} valence band (between $-10$~eV and $E_F$) would
not provide a meaningful comparison because of the superimposed
LAO and LNO valence band contributions, whose relative intensities 
strongly change for different layer thicknesses due to the finite probing depth.

In the VB spectra of the (4//4) and (10//4) SL the $e_g$ derived feature shows a
clear cut-off by the Fermi-Dirac-function at $E_F$ signalling that the samples
remain metallic at low temperature down to a LNO layer thickness of 4 uc. This
is consistent with electrical transport measurements shown in
Fig.~\ref{Fig:AngleIntSpec}~(c) and recent results from optical ellipsometry on
the same SLs. \cite{Boris11}

In comparison, the VB spectrum of the (2//2) SL displays a remarkable change near the Fermi level: 
No Fermi edge can be identified, instead both the $t_{2g}$ and $e_g$ features are
smeared out indicating a loss of quasiparticle coherence. In contrast to a
recent photoemission study of ultra-thin films\cite{Sakai13} our VB spectrum
does not show a full band gap opening, but a distinct reduction of spectral
weight at the Fermi level. Corresponding four point probe transport measurements
of the identical samples confirm an insulating phase at low temperature and a
temperature dependent MI crossover at $T \approx 150$~K driven by the reduced
dimensionality [see arrow in Fig. \ref{Fig:AngleIntSpec}~(c)]. For very thin LNO layers 
such a transition has only been observed in SL structures, but not in films. \cite{Boris11, Kumah14}

\subsection{$k$-resolved spectra}


%
\begin{figure}
\includegraphics[width = 0.48\textwidth]{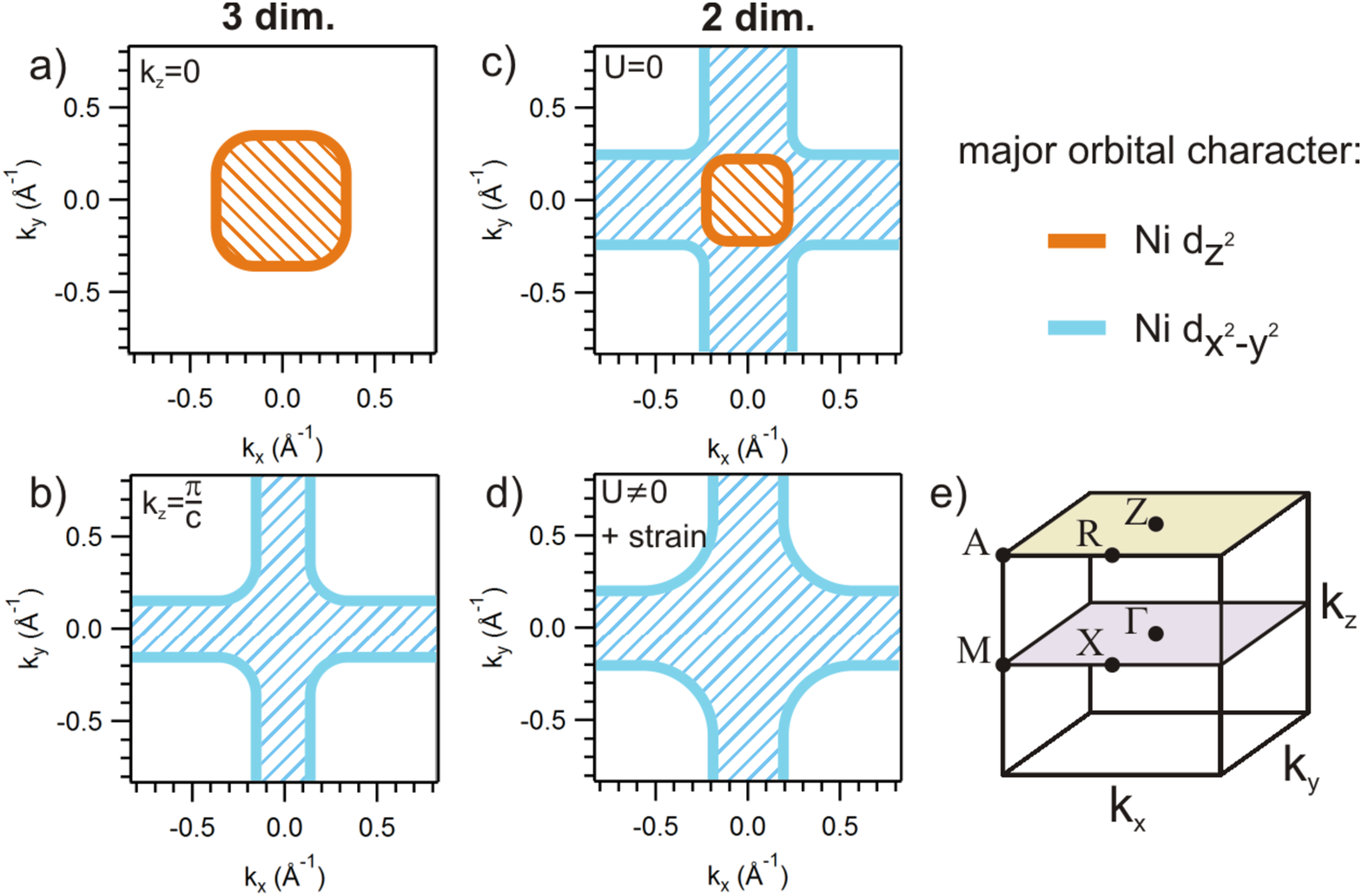}
\caption{\label{Fig:BZSketch} (Color online) Schematic cuts (solid lines) through 
the three-dimensional Fermi surface of bulk LNO at (a) $k_z=0$~{\AA}$^{-1}$ 
and (b) $k_z=\frac{\pi}{c}$ as well as the $k_z$-independent two-dimensional 
FS (c) without and (d) including correlation effects (symbolized in the figure 
by zero and non-vanishing on-site Coulomb repulsion energy $U$) and tensile 
strain according 
to Ref.~\onlinecite{Hamada93} and \onlinecite{Hansmann09}, respectively. The shaded 
areas indicate the occupied states. (e) Drawing of the cubic Brillouin zone 
including the high-symmetry points.}
\end{figure}

The angle-integrated spectra already demonstrate the pronounced effect of  
reduced LNO layer thickness on the electronic structure. In the following 
we study this in more detail by using the $k$-space resolved
spectra, in particular with respect to Fermi surface (FS) volume and topology. 
For the interpretation of the measured data it is helpful to start the discussion 
from the theoretical FS obtained by DFT calculations. Figure \ref{Fig:BZSketch} 
shows cuts through the expected three-dimensional FS of bulk LNO
parallel to the surface ($k_z=0$~{\AA}$^{-1}$ and $\frac{\pi}{c}$, $k_z$ denoting the 
wavevector component perpendicular to the surface) as well as
the $k_z$-independent two-dimensional FS of a single LNO layer (calculated without (c) and with (d) correlation effects), based on Refs. \onlinecite{Hamada93} and
\onlinecite{Hansmann09}, respectively. In the case of bulk LNO the Ni $d_{z^2}$ states form
an electron pocket at the Brillouin zone (BZ) center ($\Gamma$ point) [cf. Fig.
\ref{Fig:BZSketch}~(a)], while the Ni $d_{x^2-y^2}$ orbitals create large hole
pockets at the zone corners (A points) [cf. Fig. \ref{Fig:BZSketch}~(b)]. By
reducing the LNO layer thickness, the three-dimensional FS transforms into a
two-dimensional one, where both the electron and hole pockets are present [cf.
Fig. \ref{Fig:BZSketch}~(c)]. Hansmann {\it et al.} showed that under tensile
strain and influence of correlation effects the $d_{z^2}$ states can be lifted
above the Fermi level and fully depopulated [cf. Fig. \ref{Fig:BZSketch}~(d)].


\begin{figure}
\includegraphics[width = 0.48\textwidth]{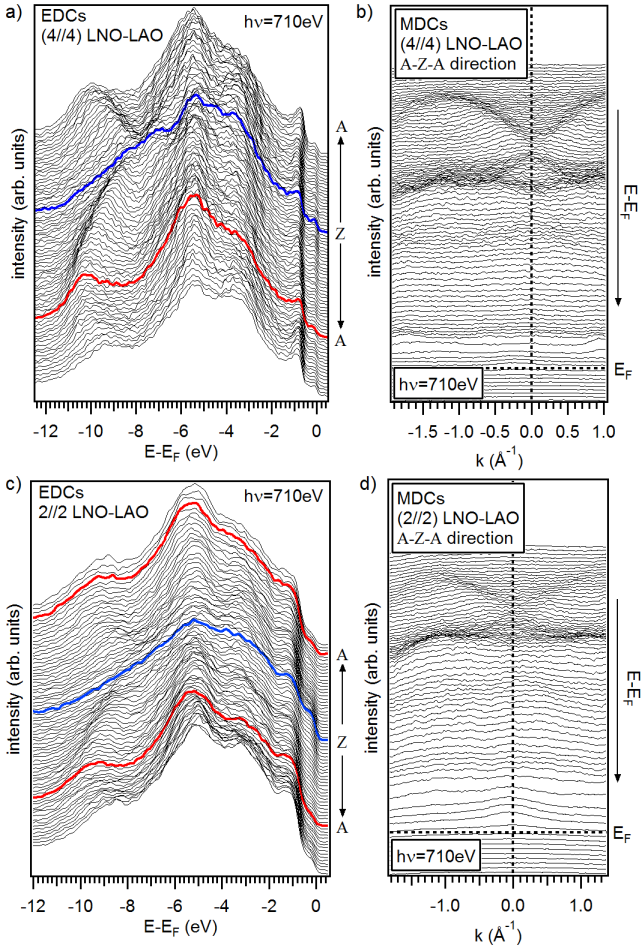}
\caption{\label{Fig:EDCsMDCs} (Color online) (a), (b) The energy and momentum 
distribution curves (EDCs and MDCs) of the (4//4) 
SL taken in A--Z--A direction show at lower energies strongly dispersive O $2p$ derived bands. 
(c), (d) These bands are also detectable in the EDCs and MDCs of the (2//2) SL measured 
in same direction. The observation of dispersive bands confirms the high crystalline quality 
of both SL and allows the identification of the high-symmetry points.}
\end{figure}

We begin our analysis of the $k$-resolved data by investigating the spectra 
taken with a photon energy of 710 eV, which corresponds to a $k$-space cut at
$k_z= \frac{\pi}{c}$ (see Appendix \ref{Sec:App2}). Here the hole pockets 
should always be observable, independent of dimensionality or possible strain- or 
correlation-induced orbital polarization. Fig.~\ref{Fig:EDCsMDCs}~(a) and (c) 
show the energy distribution curves (EDCs) and (b) and (d) the corresponding momentum
distribution curves (MDCs) of the (4//4) and (2//2) sample, respectively,
along the A--Z--A direction of the BZ. Strongly dispersive O~$2p$-derived bands are
observed in both samples at lower energies between -8 to -10~eV, indicating the 
high-quality crystalline structure of our SLs. Furthermore, the periodic band dispersion
allows a clear identification of the high symmetry points Z at
$k_\|=0$~{\AA}$^{-1}$ and A at $k_\|=\frac{\pi}{a}$. The
superposition of LAO and LNO states between -8 and -2\,eV hinders the
detection of dispersions of the O~$2p$-derived bands in this region. Near $E_F$
one can clearly identify the Ni~$3d$ states, although their intensity is much lower in
comparison to the O~$2p$-derived valence states.


\begin{figure}
\includegraphics[width = 0.42\textwidth]{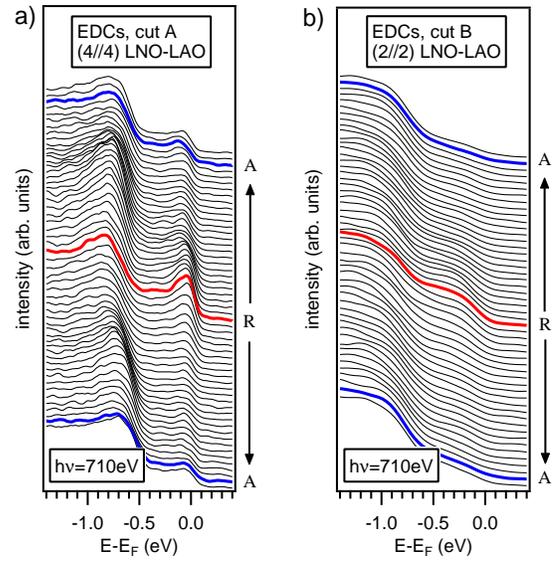}
\caption{\label{Fig:nearEF} (Color online) EDCs near $E_F$ measured along two 
cuts --- labelled A and B --- in A--R--A direction. The locations of both 
cuts are indicated in Fig.~\ref{Fig:FermiSurface}. (a) The EDCs of the (4//4) 
SL exhibit no 
dispersion in the hole pocket 
states around R within experimental resolution. (b) Both the Ni $t_{2g}$ and $e_g$ states 
of the (2//2) SL are smeared out due to the loss of QP coherence. Only a residue of 
$k$-dependent intensity can be found near R.}
\end{figure}

A more detailed view on the Ni $3d$ states near the Fermi level is given in
Fig.~\ref{Fig:nearEF}~(a) and (b), where the EDCs taken in A--R--A direction are
shown for both samples. Interestingly, a large fraction of $k$-{\it independent}
intensity at the Fermi level is found. This loss of momentum information may be
explained by non-direct transitions due to polaronic effects, as also
reported for other oxide compounds. \cite{Dessau98, Schrupp05} Nevertheless, 
there is residual $k$-dependent intensity modulation at the Fermi level in both
samples (though of different strength), consistent with the expected Fermi level crossing 
of the $d_{x^2-y^2}$ band along A--R--A [cf. Fig. \ref{Fig:BZSketch}]. The EDCs 
of the (4//4) SL [see Fig. \ref{Fig:nearEF} (a)] exhibit an intense structure 
around R which can be assigned to the 
occupied states between the hole pockets. However, while Eguchi {\it et al.} found at
this high-symmetry point a clearly dispersive band in their SX-ARPES study of a
thick LNO film under comparable experimental conditions\cite{Eguchi09}, no
dispersion is discernible in our measurement, already indicating a deviation of 
the SL
from the bulk band structure.  

In comparison to the (4//4) SL, where the Ni $t_{2g}$ and $e_{g}$ states are
well separated in energy, both features are considerably smeared out in the
(2//2) SL [see Fig. \ref{Fig:nearEF} (b)]. Since the SL is well ordered as
proven by the distinct dispersions in the O~$2p$ valence band, we attribute this behavior to the pronounced loss of QP coherence, as already discussed above.
The reduced QP coherence furthermore prevents the identification of possible
dispersive structures. An only very weak, but significant intensity modulation
around the R point may be interpreted as a remnant of the occupied
$d_{x^2-y^2}$ band states sitting on top of a largely $k$-independent incoherent background.

\subsection{X-ray photoelectron diffraction}

%
\begin{figure*}
\includegraphics[width = 0.95\textwidth]{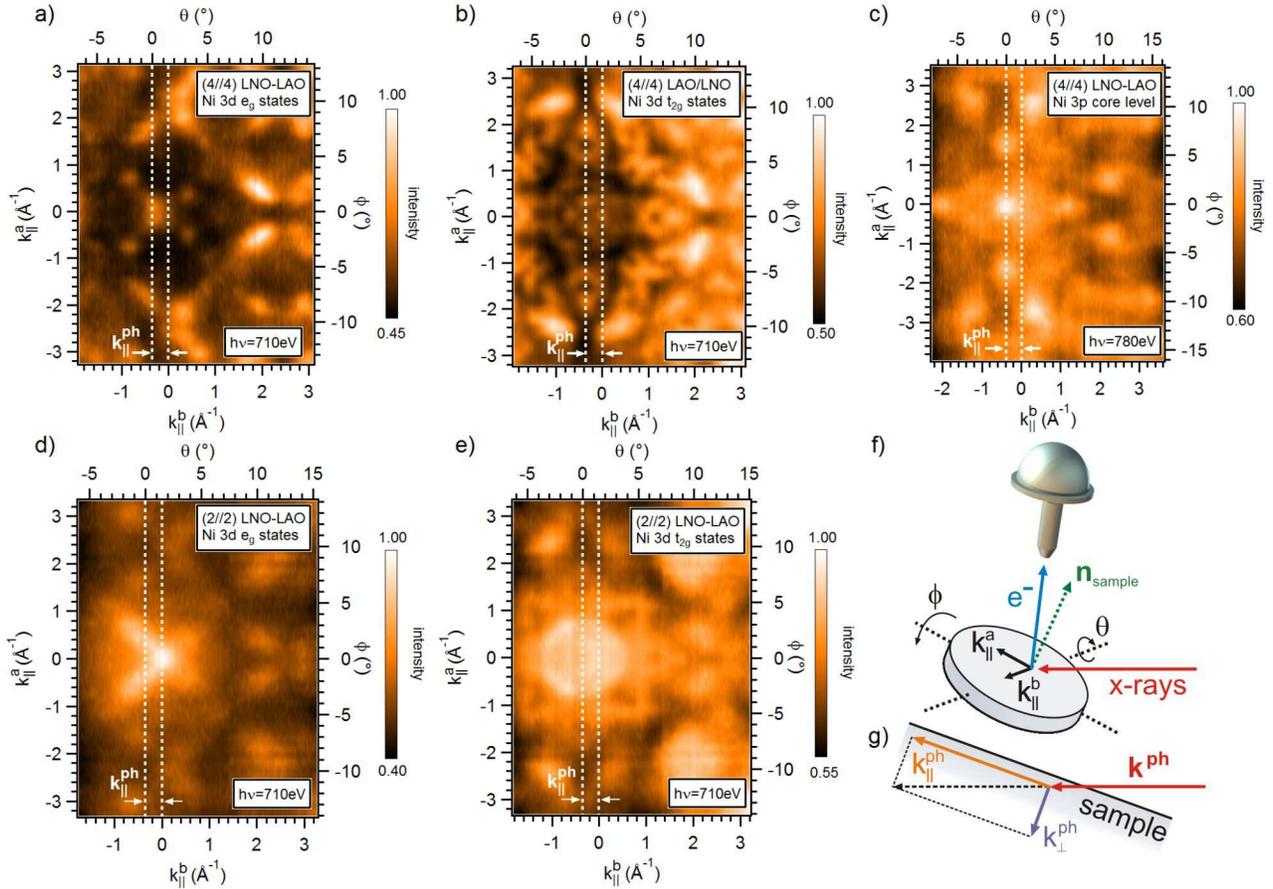} %
\caption{\label{Fig:CECs} (Color online) Angle-resolved constant energy
maps of the (4//4) and (2//2) SL measured at different binding energies. Only 
part of the full intensity range is captured by the color scale 
as is indicated for each map. (a) The map of the (4//4) SL at $E_F$ ($e_g$
states) exhibits strong intensity modulation. (b) The intensity map measured at
$-0.8$~eV ($t_{2g}$ states) displays symmetry with respect to $\theta =
0^\circ$. Thus, the observed modulations may be identified as a result of
photoelectron diffraction. (c) Angle-resolved measurements of the Ni~$3p$ core
level, taken at a similar kinetic energy as the valence band maps in (a) and
(b), show a comparable pattern. (d),(e) The angle-resolved data of the $e_g$ (d)
and $t_{2g}$ states (e) taken on the (2//2) SL provide similar results. It should be noted that the figures of the (2//2) SL were made by mirroring the data at
$\phi =0^\circ$, which does not reduce the information on the FS, since the line
$\phi =0^\circ$ corresponds to a real mirror plane of the experiment. (f)
Schematic view of the experimental geometry. 
(g) The projection of the soft x-ray photon momentum parallel and
perpendicular to the sample surface at arbitrary $\theta$ and $\phi = 0^\circ$.
}
\end{figure*}

Figures~\ref{Fig:CECs}~(a)--(e) show angle-resolved photoelectron 
distribution 
maps measured at different fixed binding energies (constant energy maps). The
experimental geometry is sketched in Fig.~\ref{Fig:CECs}~(f), where the
angles $\theta$ and $\phi$ are rotation angles around the sample axes 
as indicated and describe the direction of photoelectron emission with respect 
to the surface normal.\cite{Strocov14} By adjusting $\theta$ and $\phi$ a 
two-dimensional cut in momentum space, spanned by ${\bf k}_{\|}^a$ and 
${\bf k}_{\|}^b$, 
can be sampled (in our experiments the sample orientation was such that 
the $k_x$ and $k_y$ axes, defining the first Brillouin zone as in 
Fig.~\ref{Fig:BZSketch}, 
were rotated by 45$^{\circ}$ with respect to ${\bf k}_{\|}^a$ and 
${\bf k}_{\|}^b$). For 
the special case $\phi = 0^\circ$, the relation between 
the initial state momentum
component parallel to the sample surface, $k_{\|}^a$, and
$\theta$ is given by the following equation:
\begin{equation}
k_{\|}^a = \frac{\sqrt{2m_e}}{\hbar} \sqrt{E_\text{kin}} \sin \theta - k^\text{ph}_\|,
\end{equation}
where $m_e$ is the free electron mass and $E_\text{kin}$ the kinetic energy of
the photoelectrons. $k^\text{ph}_\|$ denotes the parallel component of the
photon momentum $p^\text{ph} = \hbar k^\text{ph} = \frac{h\nu}{c}$, which is
transferred to the emitted photoelectron [see Fig.~\ref{Fig:CECs}~(g)]. While
this contribution can be neglected in low-energy ARPES measurements 
($h\nu < 100$~eV), it has to be taken into account in SX-ARPES when converting the
angular scale to momentum scale. Note that in the given geometry a correction
only for the $\theta$-axis is sufficient, since the variation of $\phi\neq
0^\circ$ is small and thus the lateral contribution of the photon momentum is negligible.

Figure \ref{Fig:CECs}~(a) and (b) show the constant energy maps of the (4//4) SL
at the Fermi level ($e_g$ states) and at $E=-0.8$~eV ($t_{2g}$ states)
integrated over energy windows of $\pm 0.15$~eV and $\pm 0.30$~eV, respectively.
Note that only part of the full intensity range is captured by the color scale 
as is indicated for each map. Remarkable intensity modulations are observed in
the map of the $e_g$ states. A more detailed analysis may be
obtained from the angle-dependent measurements of the $t_{2g}$ states in
Fig.~\ref{Fig:CECs}~(b): The chosen integration range corresponds roughly to the
bandwidth of the fully occupied $t_{2g}$ states, and hence the spectral
weight distribution should simply reflect the momentum distribution function $n(\vec{k})$ which 
for a completely occupied band is constant $=1$. However, strong intensity modulations are
clearly observable exhibiting a geometric symmetry with respect to $\theta = 
0^\circ$ [cf. Fig.~\ref{Fig:CECs}~(f)] but none with respect to any 
high-symmetry point or line in $k$-space.
This suggests that most of the intensity modulation may be
attributed to x-ray photoelectron diffraction (XPD).\cite{Fadley90} Here, the 
excited
photoelectron is viewed as a spherical wave originating at some emitting atom
and elastically scattered off the neighboring atoms in the lattice. Interference
results in an intensity pattern which is symmetric in the 
\textit{angle coordinates} while modulations due to the electronic structure 
are symmetric in \textit{momentum space}. Note that for the high kinetic 
energies used here forward
scattering dominates the XPD process.\cite{Fadley87} The appearance of this effect
in SX-ARPES measurements has been reported also for other oxide SL
systems.\cite{Gray13}

The angular intensity distribution of the $e_g$ photoemission map 
is additionally affected by the only partial $k$-space 
occupation of the $e_g$ states, leading to the strongly reduced signal at small angles 
(with some remaining non-zero incoherent background). XPD effects are 
nonetheless well visible in the occupied part of the Brillouin zone, {\it
e.g.}, as pronounced intensity maxima at $\theta \approx 10^\circ$ (such are seen also in the $t_{2g}$ maps).

The identification of XPD as origin of the intensity modulations in
the Ni~ $3d$ states is supported by angle-resolved data of the Ni~$3p$ 
\textit{core
level} at $E = - 67$~eV [see Fig.~\ref{Fig:CECs}~(c)]. The data were taken at a
photon energy of $h\nu = 780$~eV in order to ensure the same kinetic energy, 
\textit{i.e.}, same wave length, as in the ARPES maps in Fig.~\ref{Fig:CECs}~(a) 
and (b). The Ni~$3p$ XPD pattern --- since 
core-levels are dispersionless, without intensity modulations owing to a 
momentum-dependence of the electronic structure --- shows clear 
structure-induced symmetries about $\theta=0^\circ$ 
and $\phi=0^\circ$ and contains many of the features also seen in the Ni~$3d$ maps.  

We observe similar intensity distributions in the angle-resolved maps of the 
$e_g$- and $t_{2g}$ states of the (2//2) SL, as shown in Fig. ~\ref{Fig:CECs}~(d) 
and (e), respectively. While the map of the fully occupied $t_{2g}$ band again
exhibits clear XPD-induced symmetry with respect to $\theta = 0^\circ$, the map 
of the $e_g$ states displays a superposition of $k$-dependent state
occupancy, {\it e.g.}, the X-shaped structure at $k_\|^{a,b} = 0$\,\AA$^{-1}$, and
XPD-induced modulations, {\it e.g.}, at $\theta \approx 10^\circ$.

The difference in the observed intensity modulations between both SLs at $E_F$ is
explained on the one hand by the $k$-dependent occupancy of the Ni~$e_g$ states 
(see discussion in the next section). On the other hand the different layer 
thicknesses of LAO and LNO may lead to slightly different XPD modulation patterns, 
especially in the angle-resolved intensity distribution maps of the $t_{2g}$ states.

\subsection{Fermi surfaces}

\begin{figure}
\includegraphics[width = 0.42\textwidth]{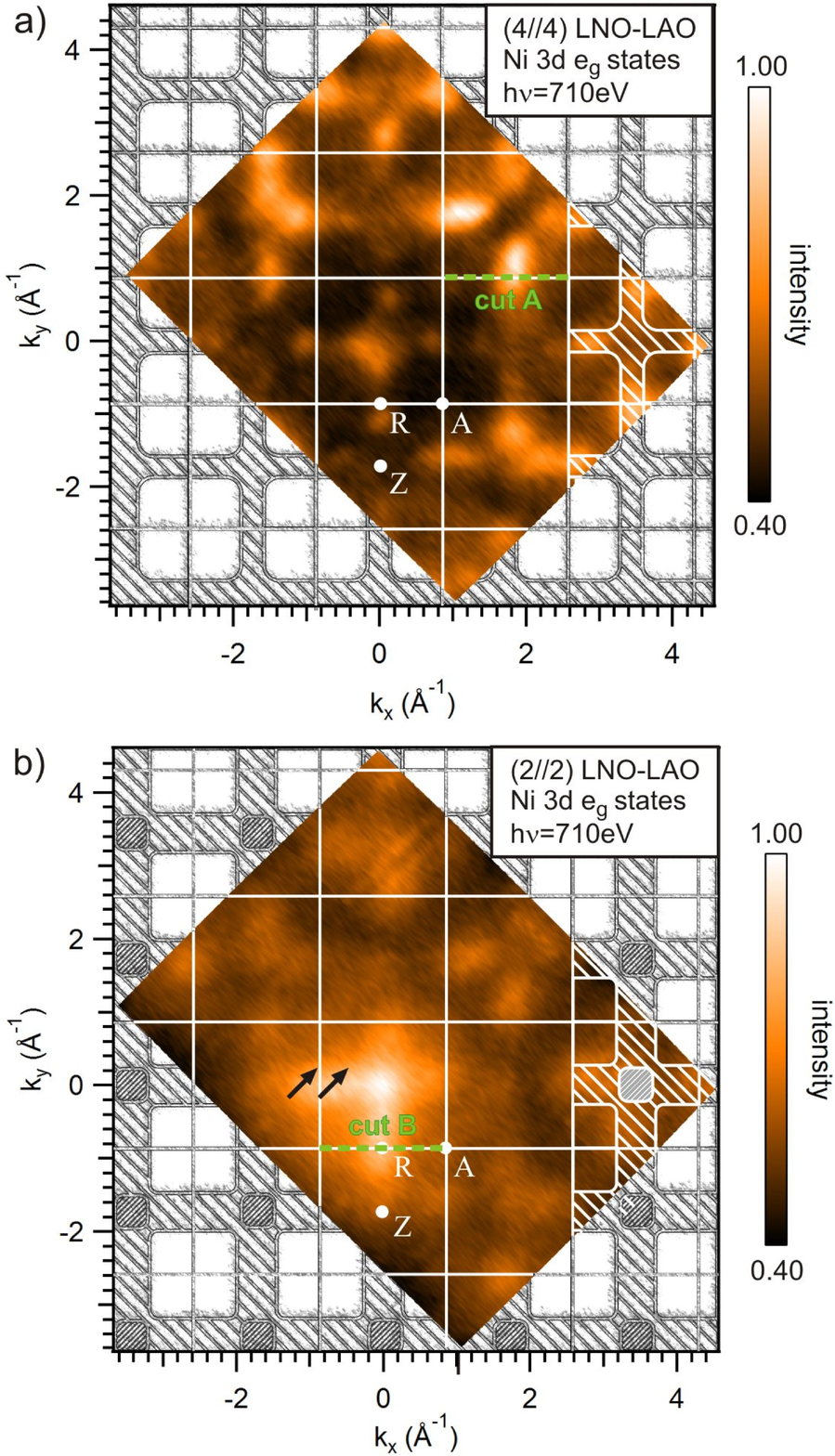}
\caption{\label{Fig:FermiSurface} (Color online) (a) The $k$-resolved intensity
distribution map of the (4//4) SL taken at $E_F$ with $k_z=\frac{\pi}{c}$
(710~eV) including several BZs shows the distinct hole pockets around the A
points, in line with band theory (shaded drawings). (b) The $k$-resolved map
of the (2//2) SL is in good agreement with the $k_z$-independent two-dimensional
FS with hole pockets around the A points and an additional electron pocket around the Z
point. The latter one is only clearly observable at $k_{x,y} = 0$~{\AA}$^{-1}$, 
presumably due to matrix element effects. The arrows indicate a possible nesting vector of the FS,
which agrees with the SDW wave vector $\mathbf{Q}_{SDW} = 2 \pi (\frac{1}{4},
\frac{1}{4}, 0)$. The lines A and B denote the $k$-space cuts probed by the EDCs
in Fig.~\ref{Fig:nearEF}. The intensity of both maps is normalized to
that at $k_{x,y} = 0$~{\AA}$^{-1}$.}
\end{figure}

\begin{figure*}
\includegraphics[width = 0.95\textwidth]{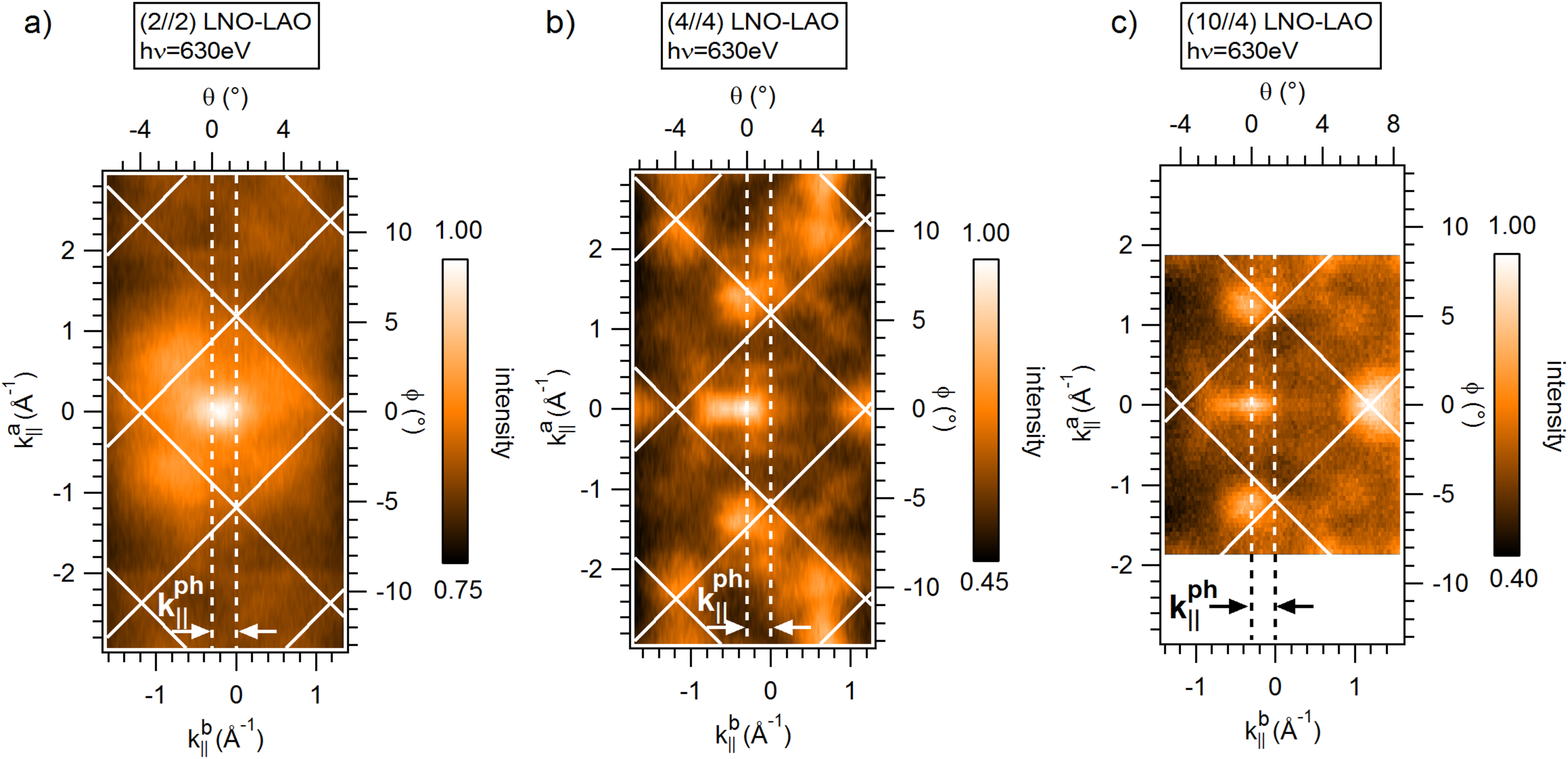}
\caption{\label{Fig:FS_630eV} (Color online) The angle-resolved intensity
distribution maps of different LNO/LAO SL at $E_F$, measured at $h\nu = 630$~eV,
which corresponds to $k_z = 0~${\AA}$^{-1}$. The white lines denote the BZ
borders. (a) The data taken on the (2//2) SL shows intensity 
at $k_\|^{a,b}=0$~{\AA}$^{-1}$, which may be assigned to the electron pocket, 
slightly interfering
with the XPD-induced intensity maximum at $\theta = 0^\circ$. The measured
intensity distribution maps of the (b) (4//4) SL and (c) (10//4) SL are
dominated by XPD effects. No electron pocket can be identified.}
\end{figure*}


After having separated out the angular XPD effects in the photoemission maps
taken at $E_F$ ($e_g$ states) we now turn to the underlying $k$-dependent
structures resulting from the electronic structure, \textit{i.e.}, the Fermi
surface of the buried LNO layers. For better identification we compare the
measured intensity distributions, provided with a momentum scale, with the
calculated Fermi surfaces in Fig.~\ref{Fig:FermiSurface}.

Starting with the (4//4) SL [Fig.~\ref{Fig:FermiSurface}~(a)], recorded at $h\nu=710$\,eV, which in the free-electron approximation for the final states corresponds to $k_z=\frac{\pi}{c}$, we find qualitatively very 
good agreement with the calculated $k$-space occupation at this very $k_z$ 
(gray shaded schematic). In particular, the hole pockets around the A points are 
clearly visible.
As already discussed above, due to the moderate resolution of the experiment no
sharp Fermi-level crossings of the dispersing bands can be observed. With the energy
integration range ($\pm 0.15$~eV) being comparable to the $e_g$ band width these
maps essentially reflect the momentum distribution function $n(\vec{k})$, {\it i.e.}, the
occupied $k$-space between the hole pockets rather than the FS contours [cf.
shaded areas in Fig.~\ref{Fig:FermiSurface}~(a)].\cite{Straub97}

The Fermi level intensity map of the (2//2) SL shown in
Fig.~\ref{Fig:FermiSurface}~(b) basically resembles the expected momentum
distribution function in the two-dimensional limit. Compared to the data of the
(4//4) SL the structures are broader in $k$-space. This can be explained
by the change of the FS from the three-dimensional to the two-dimensional
topology, where the hole pockets have to shrink in order to keep the Fermi volume
constant [cf. Fig.~\ref{Fig:BZSketch}], resulting in a broader occupied $k$-space 
region between the hole pockets. At the Z point of the first BZ
of the (2//2) SL some additional intensity is clearly visible, but much less 
so in higher-order zones, possibly due to matrix element effects. It is tempting to 
assign this intensity to the additional electron pocket in the center of the BZ
predicted for the two-dimensional limit of Fig.~\ref{Fig:BZSketch}~(c). That 
this intensity maximum is clearly centered around $k_\|^{x,y} = 0$\,\AA$^{-1}$ 
(and not $\theta = 0^\circ$, see Fig.~\ref{Fig:CECs}~(d)) supports the 
interpretation as genuine part of the band structure and rules out an XPD 
effect.

Both maps are taken at $k_z=\frac{\pi}{c}$. Since for a three-dimensional, \textit{i.e.}, bulk-like, FS the electron pocket should only be visible at the
$\Gamma$-points (corresponding to integer multiples of $\frac{2\pi}{c}$ in $k_z$ 
direction within the extended
zone scheme), as can be seen from the theoretical FS in Fig.~\ref{Fig:BZSketch}, it is interesting to look at FS cuts containing a $\Gamma$ point. The next
available $\Gamma$-point requires a photon energy of $h\nu = 630$~eV (see
Appendix~\ref{Sec:App2}). The intensity distribution map of the (2//2) SL shown
in Fig.~\ref{Fig:FS_630eV}~(a) indeed exhibits significant intensity at the
$\Gamma$ point, which may be a result of overlapping intensity derived from the
Ni~$d_{z^2}$ states at $\Gamma$ and XPD-induced modulations at $\theta =
0^\circ$. The diffuse intensity distribution towards the BZ edge may be
interpreted as residual spectral weight of the occupied states between the hole pockets, although the very low
Fermi level intensity prevents a more detailed
analysis. In any case, the ubiquity of the zone center intensity for all photon
energies (and thus its $k_z$-independence) strongly supports the two-dimensional
character of the FS in the (2//2) SL.

Figure~\ref{Fig:FS_630eV}~(b) shows the Fermi surface maps taken on the (4//4) SL
at $h\nu = 630$~eV. Due to the thicker LNO layer one would expect to slowly
recover bulk behavior with a stronger electron pocket signal at the $\Gamma$
point. However, we found no evidence for a $k$-dependent state occupancy in our
data. Rather, all intensity modulations exhibit a symmetry with respect to
$\theta = 0^\circ$, identifying them as XPD-induced structures. To verify this
result, we additionally performed angle-resolved measurements on the (10//4) SL
at the same photon energy, since at 10~uc LNO layer thickness a closer approach
to a fully established three-dimensional FS can be expected.\cite{Yoo13}
However, also in this SL no significant $k$-dependent structures are detected,
in particular, no electron pocket is found at the $\Gamma$ point [see
Fig.~\ref{Fig:FS_630eV}~(c)]. Furthermore, a $k$-space scan in $k_z$ direction
by varying the photon energy from 580 to 820~eV (not shown here) did not reveal
any signal which could be assigned to the Ni~$d_{z^2}$ pocket around
$k_\|^{a,b} = 0$~{\AA}$^{-1}$.

\section{Discussion}

The SX-ARPES data taken for $k_z=\frac{\pi}{c}$ ($h\nu = 710$~eV) provide direct 
spectroscopic evidence of the predicted Ni~$d_{x^2-y^2}$-derived hole pockets  in all 
investigated SL. However, for the electron pocket formed by Ni~$d_{z^2}$ states in the 
BZ center indications are only found in the (2//2) SL. The angle-resolved data of both the
(4//4) SL and (10//4) SL taken in the central plane of the BZ ($k_z=0$~{\AA}$^{-1}$, 
\textit{i.e.} $h\nu = 630$~eV) are strongly affected by XPD-induced intensity 
modulations and display no clear $k$-dependent state occupancy.
This is a surprising result, because a recent SX-ARPES study on \textit{thick} LNO films
has reported clear evidence of the Ni~$d_{z^2}$ states and their three-dimensional 
dispersion.\cite{Eguchi09}


It may be tempting to attribute this difference between confined LNO layers in 
SLs and bulk LNO to the strain induced by the substrate. However, a recent 
resonant reflectometry study focussing on the correlation 
between strain and orbital polarization in LNO/LAO SLs provides evidence that 
compressive strain enhances the Ni~$d_{z^2}$ bandfilling compared to the 
situation in unstrained LNO.\cite{Benckiser11,Wu13} Conversely, tensile strain 
energetically lifts the $d_{z^2}$ orbital and causes its depopulation. This 
trend is supported by DFT calculations for ultra-thin LNO films\cite{Moon12} 
and superlattices.\cite{Hansmann09,Hansmann10,Han10,Blanca_Romero11} Thus, the 
scenario that compressive strain induced by the LSAO substrate possibly leads to 
a strongly reduced Ni~$d_{z^2}$ occupancy in our samples can most likely be 
excluded.

Another explanation is a much stronger XPD effect in the SL samples compared to
bare LNO films. The data of both the (4//4)-SL and the
(10//4) SL in Fig.~\ref{Fig:FS_630eV}~(b) and (c), respectively, show nearly
identical intensity patterns, despite their different LNO layer thickness. They
share however the same thickness of the LAO capping layer, and therefore it
seems likely that most of the detected intensity modulations are caused by
photoelectron diffraction within that layer. In SX-ARPES of bulk LNO films most
of the Ni~$3d$ signal will originate from the top surface layers, whereas
in LAO-capped SLs the photoelectron wave excited from the buried LNO layers will
have to pass through the LAO overlayer and thus be subject to
\textit{additional} scattering and interference effects. The stronger XPD
intensity modulations may then distort or even obscure any underlying
$k$-dependent bandstructure information. A test of this hypothesis would require
a more systematic study of SL samples with varying LAO capping layer thickness.

A very clear and unambiguous result of our SX-ARPES study is the change
in the microscopic electronic structure from the (4//4) SL to the (2//2) SL, indicating a 
significant loss of QP coherence concomitant with the dimension-controlled MI-crossover
observed in transport. Different mechanisms have been proposed to explain the latter. For 
ultra-thin LNO films 2D Anderson localization has been discussed \cite{Scherwitzl11}, 
while other studies performed on ultra-thin films\cite{Chakhalian11} and SLs \cite{Liu11}, 
both under tensile strain, attribute the insulating phase to charge disproportionation in the 
two-dimensional ground state. Recent theoretical studies found strong evidence for
a magnetic instability, {\it i.e.} the formation of a spin density wave (SDW), 
driven by FS nesting, with a wave 
vector $\mathbf{Q}_{SDW} = 2 \pi
(\frac{1}{4}, \frac{1}{4}, \frac{1}{4}) $ as
determined from the theoretical susceptibility.\cite{Lee11, Lee11a} This 
scenario is supported by spectroscopic data on PrNiO$_3$-PrAlO$_3$ 
superlattices.\cite{Hepting14}
Experimentally, muon-spin-rotation \cite{Boris11} and resonant x-ray 
diffraction experiments \cite{Frano13} indeed observed antiferromagnetic 
ordering at the predicted wavevector, but only in the (2//2) SL,
independent of the used substrate, {\it i.e.}, the induced strain. SLs with
thicker LNO layers exhibit paramagnetic behavior, indicating that the
SDW is closely linked to the reduced dimensionality.

Our measured spectroscopic data, especially on the FS topology, provide 
further support for the SDW scenario. Despite the pronounced loss of coherent 
quasiparticle weight in the (2//2) SL, the residual intensity modulation in the 
momentum distribution function $n(\vec{k})$ reproduces a nearly quadratic shape 
of the FS hole pockets, with flat contours providing a good basis for strong FS nesting. 
Within experimental resolution the wavevector predicted for the ideal two-dimensional 
case ($\mathbf{Q}_{SDW} = 2 \pi (\frac{1}{4}, \frac{1}{4}, 0)$) is indeed 
compatible with nesting of the experimental FS [see arrows in
Fig.~\ref{Fig:FermiSurface}~(b)]. On the other hand, it may seem surprising 
that FS signal can still be observed in our low-temperature data, {\it i.e.} well 
within the SDW phase, because a nesting instability would normally open a gap 
and destroy the FS. However, as already seen in the EDCs of the (2//2) SL, there is 
no full gap opening at $E_F$, at least within our experimental resolution. This 
behavior could be attributed to an insufficient SDW stabilization due to pronounced 
order parameter fluctuations, not fully unexpected in the two-dimensional limit. 
This picture is supported by very recent magneto-resistivity measurements 
performed on similar LNO/LAO SLs.\cite{Kumah14} Hepting \textit{et 
al.}\cite{Hepting14} also find that the SDW state remains metallic, i.e. the gap 
does not encompass the entire FS. Order parameter fluctuations may also explain 
the 
suppression of quasiparticle coherence and the absence of SDW-induced 
band backfolding, in remarkable contrast to recent ARPES results on ultra-thin 
films. \cite{Yoo13}

\section{Conclusions}

In summary, we have investigated the electronic structure of 
compressively strained LNO/LAO SLs grown on LSAO substrates by angle-resolved
soft x-ray photoemission. $k$-integrated valence band spectra show a loss of quasiparticle
coherence below 3 uc LNO layer thickness. Corresponding transport measurements
exhibit a temperature dependent MI crossover in the (2//2) SL, while the (4//4) 
and (10//4)
SLs stay metallic down to low temperatures.

Although the analysis of the angle-resolved measurements is complicated by
strong XPD-induced intensity modulations, the measured angle-resolved maps
reveal a dimensional crossover of the FS from three-dimensional in the (4//4)
SL to two-dimensional behavior in the (2//2) SL. By comparing the maps with results
from DFT calculations, the Ni $d_{x^2-y^2}$ states, which form the hole states
around the A points, are clearly identified in the FS of all measured SLs. Evidence for
the electron pocket derived from Ni~$d_{z^2}$ orbital is found only in the
(2//2) SL, but could not be observed for SLs with thicker LNO and LAO layers. 
We attribute this to pronounced XPD effects in the 
LAO overlayer, which interfere with or even obscure $k$-dependent 
Fermi surface information

The measured FS topology, in particular the shape of hole pockets in the (2//2) SL, support
FS nesting. This is consistent with the scenario of a dimensionality-induced SDW-instability, 
with the nesting properties of our experimental FS in excellent agreement with the reported 
SDW wavevector. Thus, our results strongly support magnetic ordering in the
two-dimensional ground state of ultra-thin LNO layers embedded in a SL. In line
with other experimental as well as theoretical studies, no indication of
correlation-induced FS modifications was found for the SLs under
compressive strain. Further studies, particularly on LNO/LAO SLs under \textit{tensile}
strain, are needed to investigate the tuneability of the electronic structure by 
strain {\it and} correlation effects in the two-dimensional limit.


\appendix

\section{Valence band offset of the LaAlO$_3$ overlayer}
\label{Sec:App1}
\begin{figure}
\includegraphics[width = 0.30\textwidth]{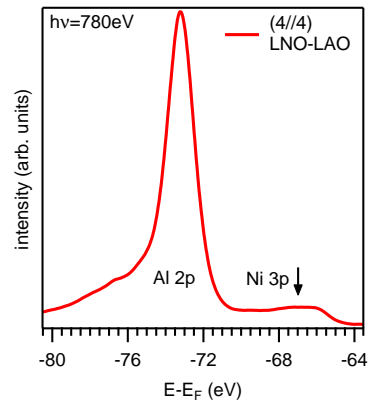}
\caption{\label{Fig:Al2pNi3p} (Color online) Core-level spectrum including the
Al~$2p$ and Ni~$3p$ line taken at $h\nu=780$~eV. From the position of the
Al~$2p$ line the valence band offset of the LAO overlayer is determined. The
arrow indicates the binding energy used for the angle-resolved core-level map
in Fig.~\ref{Fig:CECs}~(c).}
\end{figure}

In the SLs the LNO layer is buried below a LAO capping layer. Thus, the measured
valence band is a superposition of the LAO and LNO valence states. Only, if the
valence band maximum of the insulating LAO,
$E_{\text{VBM}}^{\text{LNO/LAO}}$, is far enough below the Fermi level $E_F$, 
the observed spectral weight near $E_F$ can unambiguously be assigned to the Ni~$3d$ 
states in the LNO layers. $E_{\text{VBM}}^{\text{LNO/LAO}}$ is determined by analyzing the position of a
suitable core level from LAO, $e.g$, Al~$2p$, and comparing its position with a
reference system containing LAO, where core-level position and valence band
onset are well known. Here, we use as reference the well-studied
LaAlO$_3$/SrTiO$_3$ heterostructure with $E_{\text{Al\,}2p}^{\text{LAO/STO}}
\approx -74.35$~eV and $E_{\text{VBM}}^{\text{LAO/STO}} \approx -3.1$~eV (Ref.
\onlinecite{Berner13a}). The energy position of the Al~$2p$ core level in the
LNO/LAO SLs is found to be $E_{\text{Al\,}2p} = -73.20$~eV [see Fig.~\ref{Fig:Al2pNi3p}],
independent of the LAO and LNO layer thickness. From the obvious relationship
\begin{equation}
E_{\text{VBM}}^{\text{LNO/LAO}} = E_{\text{Al\,}2p}^{\text{LNO/LAO}} - E_{\text{Al\,}2p}^{\text{LAO/STO}} + E_{\text{VBM}}^{\text{LAO/STO}}
\end{equation}
we determine the position of the valence band maximum in the LAO overlayer as 
$E_\text{VBM}^{\text{LNO/LAO}} = -1.95 \pm 0.1$~eV. Consequently, any spectral 
weight measured between $E_F$ and $\approx -2$~eV can only result from the 
Ni~$3d$ states in the LNO layers.

\section{$k_z$-dependence for bulk LaNiO$_3$}
\label{Sec:App2}

The relationship between the kinetic energy of the photoelectrons and the
momentum perpendicular to the sample surface is described by the following
equation including the non-negligible photon momentum component
$k_\perp^{\text{ph}}$ [see Fig.~\ref{Fig:CECs}~(g)]:
\begin{equation}
 k_z = \sqrt{2 m_e / \hbar^2} \left( V_0 + E_k \cos^2\theta\right)^{1/2} - k_\perp^{\text{ph}},
\end{equation}
where $m_e$ is the free electron mass, $V_0$ the inner potential, $E_k$ the
kinetic energy, and $\theta$ the emission angle. For thick LNO films a value of 
$V_0 = 10$~eV has been reported.\cite{Eguchi09} We do not expect a large 
variation of this phenomenological parameter for our LNO/LAO SLs.
Thus, by using the lattice constant perpendicular to the surface $c=3.840${\AA}
(Ref.~\onlinecite{Wu13}) and taking the experimental geometry into account, the 
photon energies required for probing Fermi level states at the center ($k_z =0$~{\AA}$^{-1}$) and
the edge of the BZ ($k_z = \frac{\pi}{c}$) perpendicular to the surface are 
$h\nu=630$~eV and $h\nu=710$~eV, respectively.

\begin{acknowledgments}
We thank C.S. Fadley and A.X. Gray for fruitful discussions. This work was
supported by the Deutsche Forschungsgemeinschaft (FOR 1162) and the German
Federal Ministry for Education and Research (05K10WW1).
\end{acknowledgments}


%

\end{document}